\documentclass{vgtc}                          
\vgtcinsertpkg

\ifpdf
  \pdfoutput=1\relax                   
  \usepackage{graphicx}                
  \DeclareGraphicsExtensions{.pdf,.png,.jpg,.jpeg} 
\else
  \usepackage{graphicx}                
  \DeclareGraphicsExtensions{.eps}     
\fi%

\graphicspath{{figures/}{pictures/}{images/}{gfx/}{./}} 

\usepackage{microtype}                 
\PassOptionsToPackage{warn}{textcomp}  
\usepackage{textcomp}                  
\usepackage{mathptmx}                  
\usepackage{times}                     
\usepackage{cite}                      
\usepackage{tabu}                      
\usepackage{booktabs}                  
\usepackage[autostyle=true]{csquotes}
\usepackage{amsmath}
\usepackage{acronym}
\usepackage[dvipsnames]{xcolor}
\usepackage{subcaption}
\usepackage{wrapfig}

\onlineid{0}
\vgtccategory{Position}

\title{Speculative Execution for Guided Visual Analytics}

\newcommand\skipand{\end{tabular}\\[1.5ex]\begin{tabular}[t]{c}}

\author{Fabian Sperrle\thanks{e-mail: fabian.sperrle@uni-konstanz.de}\\ %
        \scriptsize University of Konstanz %
\and J\"urgen Bernard\thanks{juergen.bernard@gris.tu-darmstadt.de}\\ %
     \scriptsize Technische Universit\"at Darmstadt %
\and Michael Sedlmair\thanks{michael.sedlmair@visus.uni-stuttgart.de}\\ %
     \scriptsize University of Stuttgart%
    \skipand 
Daniel Keim\thanks{e-mail: daniel.keim@uni-konstanz.de}\\ %
     \scriptsize University of Konstanz %
\and Mennatallah El-Assady\thanks{e-mail: mennatallah.el-assady@uni-konstanz.de}\\ %
     \scriptsize University of Konstanz \\
     \scriptsize University of Ontario Institute of Technology} 
     
\teaser{
 \centering
 \includegraphics[width=\linewidth]{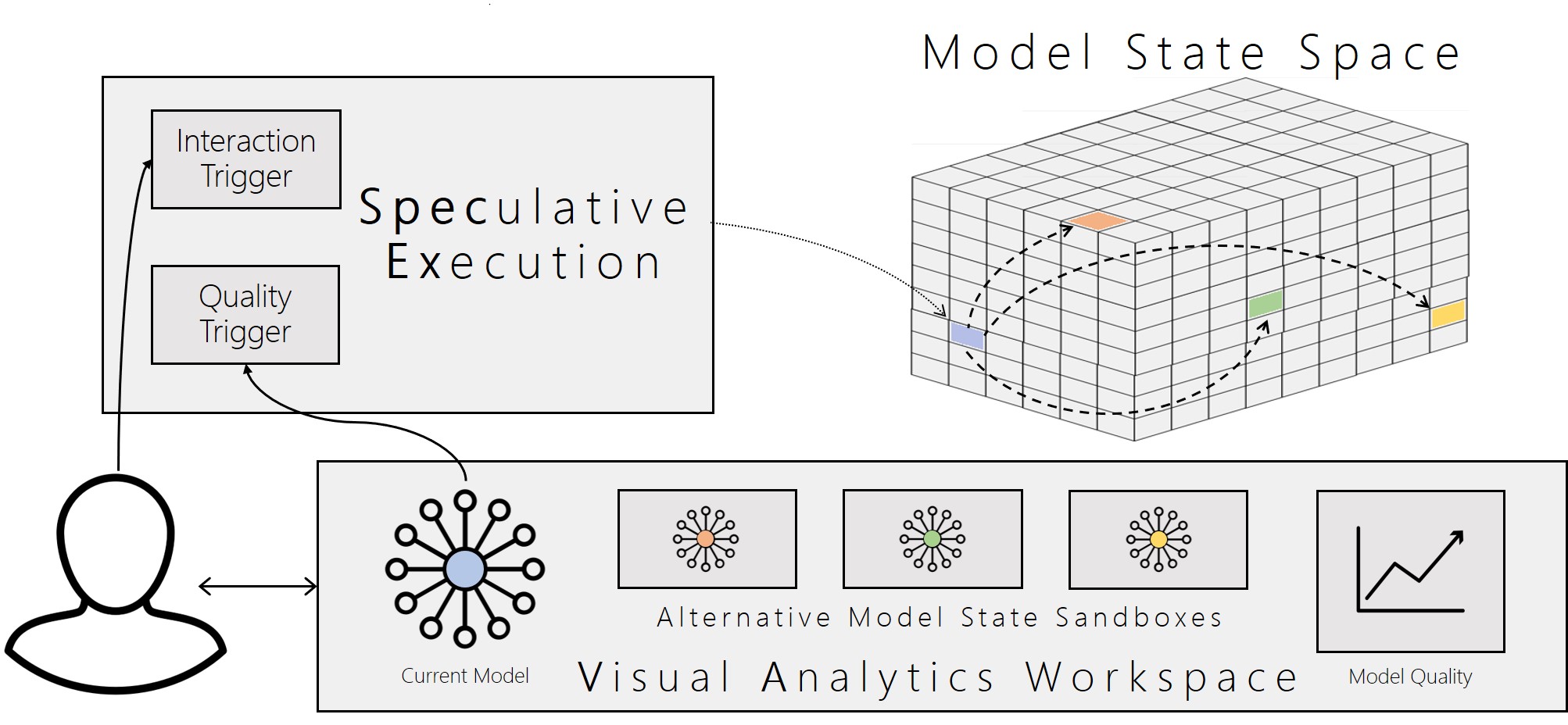}
 \caption{\acl{se} for \acl{va}. A model optimization process defines a path through the model state space (shown with only three dimensions here for simplicity). \acl{se} proactively and automatically computes alternative model states in sandboxes and presents them using delta-visualizations, comparing them to the current model. \acl{se} can be triggered either through user interaction or model quality metrics. }
}

\abstract{We propose the concept of Speculative Execution for Visual Analytics and discuss its effectiveness for model exploration and optimization.  Speculative Execution enables the automatic generation of  alternative, competing model configurations that do not alter the current model state unless explicitly confirmed by the user. These alternatives are computed based on either user interactions or model quality measures and can be explored using delta-visualizations. By automatically proposing modeling alternatives, systems employing Speculative Execution can shorten the gap between users and models, reduce the confirmation bias and speed up optimization processes. In this paper, we have assembled five application scenarios showcasing the potential of Speculative Execution, as well as a potential for further research. 
}

\CCScatlist{
  \CCScatTwelve{Human-centered computing}{Visual Analytics}{Speculative Execution}{SpecEx};
}

\begin{document}

\newacro{va}[VA]{Visual Analytics}
\newacro{se}[SpecEx]{Speculative Execution}

\firstsection{Introduction}
\maketitle

In many disciplines, domain experts have to create and validate multiple possible solutions to a current situation. With the identification of a best possible outcome, a solution can be pursued.
In politics, simulations and anticipations have been an integral part of successful campaigns for a long time: politicians and their advisers prepare for different, likely outcomes of an event or a meeting.  This allows them to be prepared for different situations that might come up and ensures that they can react in a fast, yet precise way. Similarly, politicians and involved policy analysts create, analyze, and compare alternative solutions before a policy is to be implemented~\cite{Ruppert2013}. Although most of the potential solutions will not be needed in the end, their preparation is imperative to guarantee a systematic and timely decision process.
In the world of computers, CPUs, for instance, pre-compute conditional code blocks before the outcome of the condition is known in order to avoid waiting and doing nothing. Similarly, multiple pages are fetched from disk or from memory, even when only one has been requested. All of these examples employ \acl{se}, the principle of preparing or precomputing the result of a task at a time when it is not needed, but its calculation is easier or cheaper thanks to synergy effects or idling resources. 

We propose to apply \acl{se} as a concept for \acl{va}, as well. There, it can simplify user interactions and propose parameter changes. It is inspired by the human-in-the-loop concept that has become increasingly popular in \acl{va} over the last years. It integrates human decision-making into the analysis process to obtain results that are based on semantic understanding and fit the user's expected mental model. More recently, Endert et al. proposed a pattern named ``the human is the loop''~\cite{endert2014human}. They call for new directions of Visual Analytics that focus on recognizing the user's work process and ``seamlessly fitting analytics into that existing interactive process''~\cite{endert2014human}. Speculative execution picks up this idea and proposes that systems learn the users' goals from their interactions and provide optimizations that help to reach this goal faster.  Additionally, Endert et al. also argue that ``implicit steering is perhaps the ultimate form of in-context input''. As Speculative Execution is aimed at understanding the users' intentions and executing them in the model, it can remove the need for unnecessary, explicit interactions, which are instead performed automatically by the system.  

In addition to reducing the gap between users and machine learning models, \acl{se} can also  be helpful for model optimization and model understanding. As a consequence of the increasing complexity of (machine learning) models, interactions with model visualizations are important to foster model-understanding and trust-building~\cite{Liu2017TowardsPerspective}. However, these interactions are not always straight-forward. For example, Lee et al. have found that even ``seemingly small changes can have unexpectedly large consequences''~\cite{Lee2017TheModels} on the output of the popular topic model LDA~\cite{Blei2003LatentAllocation}. This often leads to users being cautious when interacting with models in fear of ``breaking something''~\cite{El-Assady2018ThreadReconstructor}. To avoid potentially worsening the current model, Speculative Execution provides isolated computation environments. In those, changes can be applied speculatively.  In combination with a delta-visualization between two model states, this allows users to preview the model changes introduced by their interaction in a focused manner. 

We deem \ac{se} particularly beneficial for the typical \ac{va} scenarios. First, it provides the means for effective model optimization and refinement towards the users' tasks and data. Second, it can help prevent confirmation bias by showing potential modeling alternatives. Finally, it is well-suited for mixed-initiative  user-guidance. However, \ac{se} has not yet been formally introduced to \ac{va}.
We demonstrate the applicability of \ac{se} for guided \acl{va} in five usage scenarios in \autoref{sec:application_scenarios}. These scenarios also highlight that \ac{se} integrates well with existing \ac{va} concepts that will be introduced in \autoref{sec:related_work}.

The contributions of this paper are (1) the introduction of the concept of \acl{se} to \acl{va}; (2) an illustration of the value of \acl{se} in five usage scenarios; (3) design considerations and an implementation model for \acl{se}; (4) the identification of open research questions for efficient application of \acl{se}.
 
\section{Background}
\label{sec:related_work}
\ac{se} as a concept is broadly applicable in many \ac{va} scenarios. We discuss related \ac{va} techniques here, and present derived usage scenarios in \autoref{sec:application_scenarios}. 

Van den Elzen and van Wijk have introduced a visual exploration technique called ``Small Multiples, Large Singles''~\cite{van2013small} that is similar to Speculative Execution. Starting from a ``large single'' visualization a small multiples visualization shows alternative models, model parameters, visual mappings, or visualization techniques to the user. Following an alternating sequence of large singles and small multiples, users can either use the tool to explore the data space or to adjust both the model and visualization for their use case. While the approach focuses on the navigation-support provided with the small multiples (e.g., by trial-and-error), the \ac{se} concept also incorporates the suggestion of next step that may be meaningful to succeed in some task.

\ac{se} integrates well with the concept of provenance tracking. Systems like AVOCADO~\cite{stitz2016avocado}, VisTrails~\cite{callahan2006vistrails} or SenseMap~\cite{nguyen2016sensemap} focus on visualizing the provenance of results during or after an analysis session. Some systems also give users the possibility to revert to a previously seen model configuration. \ac{se} enables a more straightforward comparison of two model states, as they can both be instantiated in isolated environments at the same time. Combined with a delta-visualization this provides a powerful tool for understanding how specific interactions have influenced the model building process. 

If Speculative Execution is based on quality metrics instead of user interactions, it becomes an alternative to Visual Parameter Space Analysis, e.g., conducted by Sedlmair et al.~\cite{sedlmair2014visual}. In this case, sandboxes would be created with different parameter configurations following the speculation dimensions presented in \autoref{sec:speculation_dimensions}. However, as \ac{se} is an interactive, mixed-initiative approach, the exploration of the search space and computation of alternative models can be more restricted to those in which users are interested. Also, this interactivity allows Speculative Execution to modify the parameter sampling methods, focusing on parameter ranges that seem to be relevant to the users in a given analysis session. 

The concept of \ac{se} pairs well with progressive computation.
As \ac{se} observes user interactions with the system (see \autoref{sec:speculation_levels}), systems can observe which sandboxes users are most interested in, and progressively refine their models. This allows initial sandbox models to be computed with a lower degree of detail, enabling the computation of more sandboxes with the same resources. The necessary details can then be computed once they are necessary.

\section{Speculative Execution as a VA Concept}
We first define \emph{\acl{se}} and \emph{Computational Sandboxes}, as well as the \emph{Model Search Space}. We then combine the introduced concepts into an \emph{implementation model} that can be applied to existing \acl{va} systems. 

\subsection{Definitions}
All three of the important terms have previously been used in different areas of computer science. We present related definitions that are tailored towards the use of the terms in \acl{va}.

\paragraph{Speculative Execution}
Typically, \emph{\acl{se}} describes a set of CPU optimization techniques like branch prediction~\cite{zilles2001execution} or data prefetching~\cite{gonzalez1997speculative} that can improve their performance.
In the context of \acl{va}, we describe \ac{se} as the proactive computation of alternative, competing model states that are isolated from the current model state and do not influence it. On the one hand, such a proactive computation can be triggered by user interaction. In this case, the computation completes the interaction or performs similar operations. More detailed distinctions will be made in \autoref{sec:speculation_levels}. On the other hand, the computations can have the goal of model optimization. They are then typically triggered by model quality metrics and can, for example, be used to explore different parameter settings. Combining these two aspects, we define \ac{se} as follows:
\blockquote{\emph{\acl{se} describes the proactive, near real-time computation of competing model alternatives that do not influence the current model state, explores the model state search space, and is triggered by either interaction or quality measures.}}

Comparing the differences between two competing model states, users can decide whether any \ac{se} was useful or not. If yes, they can accept the proposed sandbox, or else reject it. Using these two simple operations, users can steer how the model state space is searched. Over time, \ac{se} can learn from this user interaction and rank proposed sandboxes higher if they conform to similar schemas as sandboxes that have previously been accepted by the user. This improves the quality of suggestions and facilitates the intended tasks of model optimization and exploration further. Imagining model optimization as a ``walk'' through the model state space, \ac{se} allows users to ``look to the left and right of the path''. This differentiates \ac{se} from ``normal'' human-in-the-loop \acl{va} that incorporates the user without necessarily focusing on exploration or proposing modeling alternatives.

Sandboxes---isolated containers holding models--- and the model state search space---the theoretical, high-dimensional space of possible model configurations---will be introduced and defined in the following paragraphs.

\paragraph{Sandboxes}
\label{sandboxes}
In general, sandboxes are environments with limited connections to their respective outside world, enabling a separated and encapsulated computation.
In computer science, the term sandbox is often used to describe shielded environments or local working copies that are used to execute some task.
Application areas include virtual machines, software installation facilities, or security-related environments.
Sandboxes in visual analytics have often been used in collaborative or multi-device environments to show all users different and tailored views of the data~\cite{jentner2016dynamite}. ``Sandbox'' is also the name of a visual sensemaking system introduced by Wright et al.~\cite{wright2006sandbox}. Additionally, ``sandboxes'' is a term often used in relation with computer- and system-security: browsers use JavaScript sandboxes~\cite{dewald2010adsandbox, ingram2012treehouse} and systems move suspect executables to sandboxes~\cite{Yee2009Sandbox}.

Building upon these characterizations and examples, we adopt the notion of sandboxes and describe them as \emph{isolated computation environments} that can be initialized with any model state. The result of any sandbox computation should always be a valid model state, such that any model can be replaced with a sandbox derived from it easily. These sandboxes can help to explore different, alternative hypotheses, which analysts often do to avoid confirmation bias~\cite{heuer1999psychology}.

\paragraph{Search Space}
\label{searchSpace}
The search space for \acl{se} is defined by all possible model states, given by the input data and all parameters. \ac{se} defines any number of these inputs as \emph{speculation dimensions} that can be modified before starting a proactive computation. 
As a result, the \acl{se} Space is significantly smaller than the entire model state space, as it does typically not make sense to change all variables before launching a computation. 
An elaboration on the different potential sizes of the search space can be found in \autoref{sec:search_space_size}.

\subsection{Implementation Model for \ac{va} with \ac{se}}
\begin{figure}[t]
    \centering
    \includegraphics[width=\linewidth]{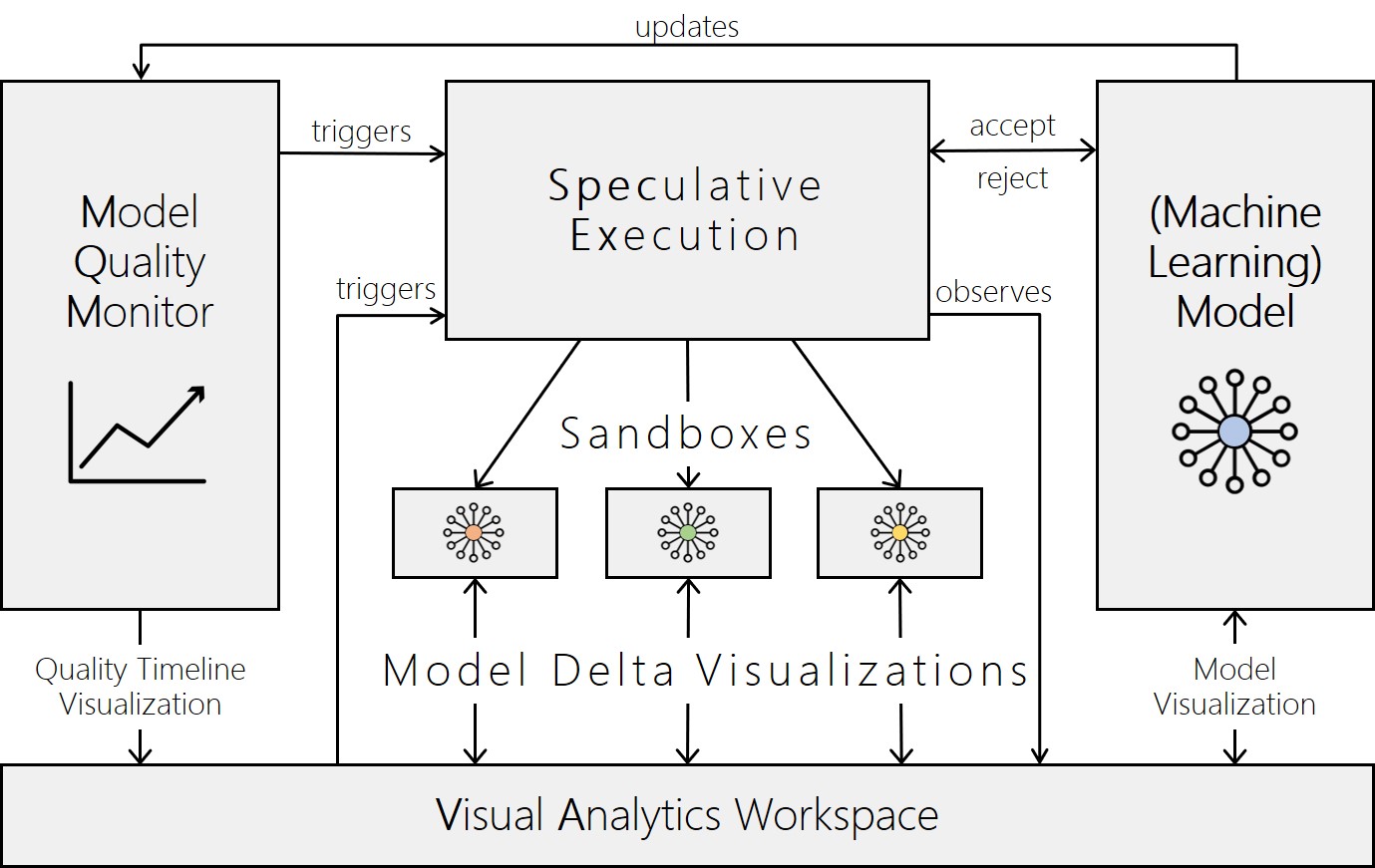}
    \caption{Implementation architecture for \acl{se} in \acl{va}. The \emph{Speculative Execution} and \emph{Model Quality Monitor} components can be retrofitted to existing \ac{va} systems, together with one or multiple \emph{Model Delta Visualizations}. }
    \label{fig:specex_workspace}
\end{figure}

Having defined the concepts of \acl{se}, we now describe how it can be integrated into an existing \ac{va} workspace. The result of such an integration is shown in \autoref{fig:specex_workspace}. The central component is the \ac{va} workspace itself with visualizations of both the model and, if available, a set of quality metrics computed by the \emph{Model Quality Monitor}. The \emph{Speculative Execution Component} can create model sandboxes and instantiate them with copies of the current model. These sandboxes can either be visualized individually or in a delta-view comparing them to a different model state; typically the ``original'' model from which the sandbox was started. The \emph{Speculative Execution Component} also constantly monitors the \ac{va} workspace and tracks the users interactions, searching for patterns and trying to determine the users intentions. The possible outcomes of this process will be described in \autoref{sec:speculation_levels}. Thus, any sandbox can either be triggered by observed user interactions or, more simply, by the \emph{Model Quality Monitor}. This component combines multiple metrics indicating the quality of the model, and employs different strategies for combining these metric values into a ``trigger-decision''. This generic implementation model can be taken into account when designing new \acl{va} systems, or be retrofitted to existing \ac{va} systems thanks to its modularity. 
It follows the architecture for ``human centered machine learning'' presented by Sacha et al.~\cite{SACHA2017164}. Speculative interaction ties into the \emph{Validation \& Interaction} stage of their proposed framework.

\ac{se} can reuse existing techniques to determine the users intentions~\cite{brown2014waldo, Dou2009Recovering}, and employ them in sandboxes. The ranking and proposition of sandboxes to the user can be incrementally improved throughout the analysis session. For example, a higher weight can be given to those sandboxes that have been created according to a schema that has often been accepted by the user in the past. However, such adaptations to the weighting scheme have to be carefully considered to avoid creating and confirming biases in the user's mental model.

\section{Usage Scenarios}
\label{sec:application_scenarios}
{
\setlength{\columnsep}{1em}%
\setlength{\intextsep}{0.5em}%
To illustrate the idea of \acl{se} in \acl{va}, we now discuss five usage scenarios in which \ac{se} was either already applied successfully, or that would build a valuable basis for an extension towards \ac{se}. 
The first scenario reports the results of the initial implementation of \ac{se} for topic modeling. The second and third scenario highlight the use of \ac{se} for implicit steering and user guidance, respectively. 
Scenario four exemplifies the value of \ac{se} for cooperative \ac{va}, and scenario five shows that \ac{se} can also be applied to visualization and is not limited to \ac{va}. Except for scenario one, these usage scenarios are hypothetical and have not been implemented yet.
The small figures under each usage scenario repeat the implementation architecture model from \autoref{fig:specex_workspace}. The individual components that that are ``active'' in each of the scenarios have been coloured in blue, highlighting the versatility of \ac{se}. Different amounts of sandboxes have been coloured throughout the examples and represent whether a small, medium or large number of sandboxes are expected to be  computed.

\subsection{Model Optimization}
\label{sec:model_optimization}
\begin{figure}
    \centering
    \includegraphics[width=\linewidth]{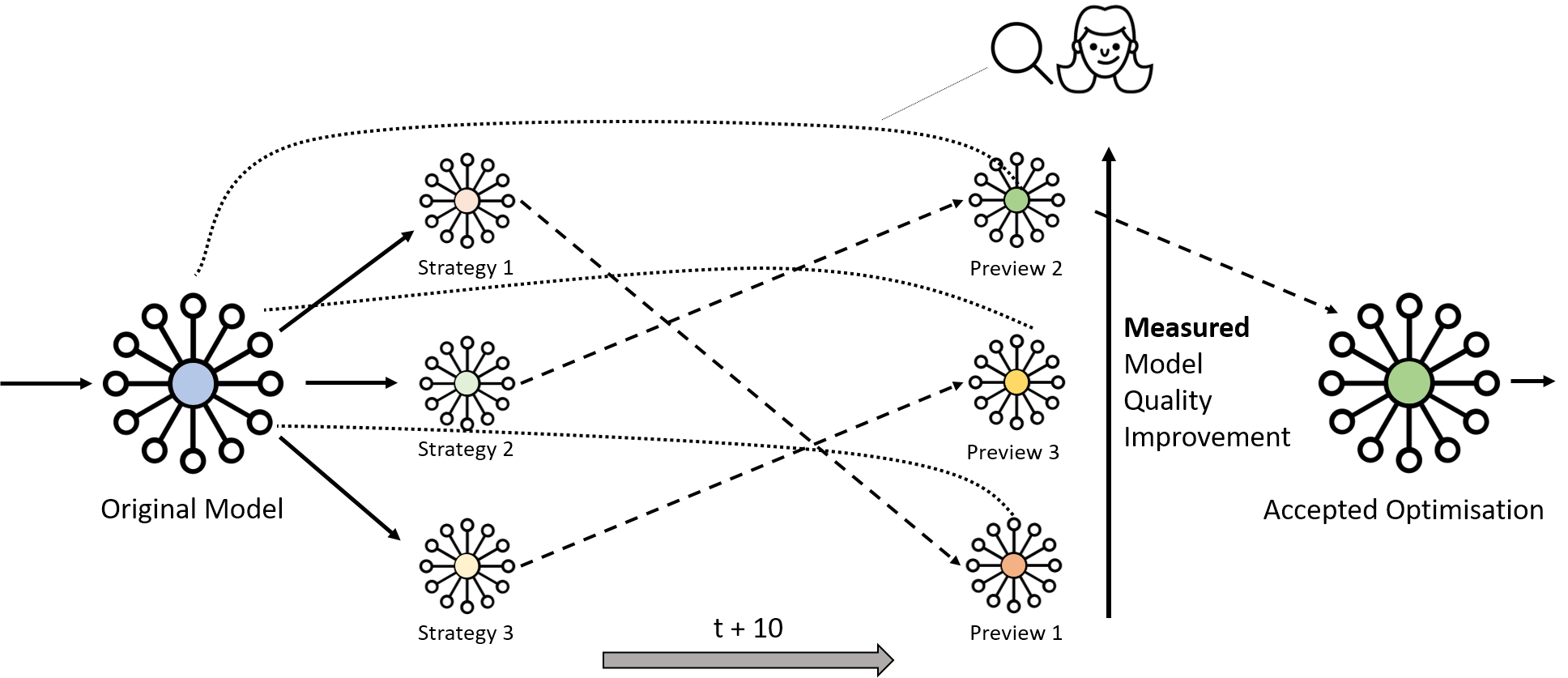}
    \caption{Example of two-dimensional SpecEx over time and different optimization strategies. Three optimization strategies for an incremental topic model are triggered in sandboxes and forecast over the next ten document inserts. The resulting models are sorted according to their quality, and presented to users as a diff with the current model. Applying their domain knowledge, users select the best model to continue the process with. }
    \label{fig:specex_example1}
\end{figure}
Topic modeling is a popular technique to segment a text corpus into thematically related clusters. Consequently, refining the results of topic models and adapting them to a particular set of users, data, and tasks, is an active area of research~\cite{El-Assady2018Progressive}. In our recent work, we examined the optimization of the Incremental Hierarchical Topic Model (IHTM) using \ac{se}~\cite{El-Assady2018SpecEx}.
\begin{wrapfigure}[7]{r}{0.4\columnwidth} 
    \vspace{-0.25em}
    \includegraphics[width=\linewidth]{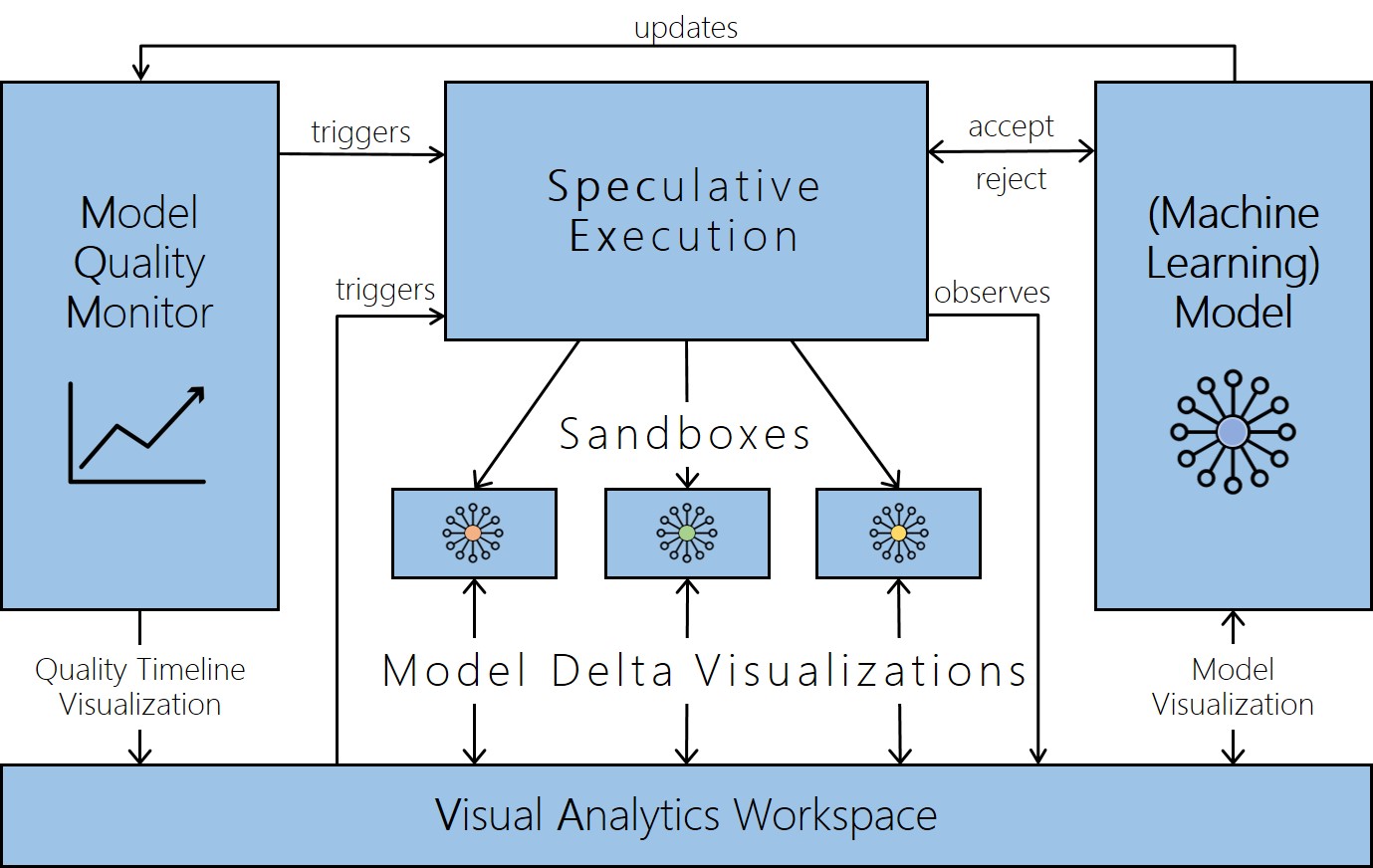}
\end{wrapfigure}
This model represents topics as a tree---with leaf nodes being documents, and inner nodes representing hierarchically ordered topics. This data structure can be explicitly manipulated through various optimization strategies that, for example, split or merge topics, remove outliers, or compact the topic tree. In combination with typical topic model quality metrics like the number and size of topics, their coherence or pointwise mutual information, a system can speculatively start the available optimization strategies in individual sandboxes whenever any (combination) of the metrics declines. Once the sandbox computations are complete, the results can be ranked according to their measured quality and be presented to the user. We have implemented and successfully tested such a system in our previous work~\cite{El-Assady2018SpecEx}.
The topic trees of two topic models are merged into a single tree, and added, moved and removed topics and documents are highlighted to help users see differences between the two models and select a sandbox to continue the computation with.  This workflow is outlined in \autoref{fig:specex_example1}.

As the user observes the model building process of the IHTM on the ``20news'' dataset, a document from the ``atheism'' group is added to the ``christian'' topic, making it too broad and leading to \acl{se}. The user inspects the proposed optimization leading to the highest measured quality improvement: merging the two topics ``mideast'' and ``baseball''. While this merge optimizes the quality metrics, the user quickly rejects it, employing their semantic understanding. Instead, they select a different sandbox in which the erroneously combined ``car'' and ``gun'' topics have been split. 

Using \ac{se} the analysis system was able to integrate the user into the model steering process seamlessly. Instead of defining must-link and cannot-link constraints to optimize topics, the user could select from a set of prepared optimizations. Especially in topic models, where often subtle semantic differences decide over the quality or even correctness of a topic attribution, \ac{se} can help to achieve good results by exploring a wider area of the search space instead of only a single model.

\subsection{Implicit Steering}
Our second example addresses the task of labeling datasets 
\begin{wrapfigure}[7]{r}{0.4\columnwidth} 
    \vspace{-0.25em}
    \includegraphics[width=\linewidth]{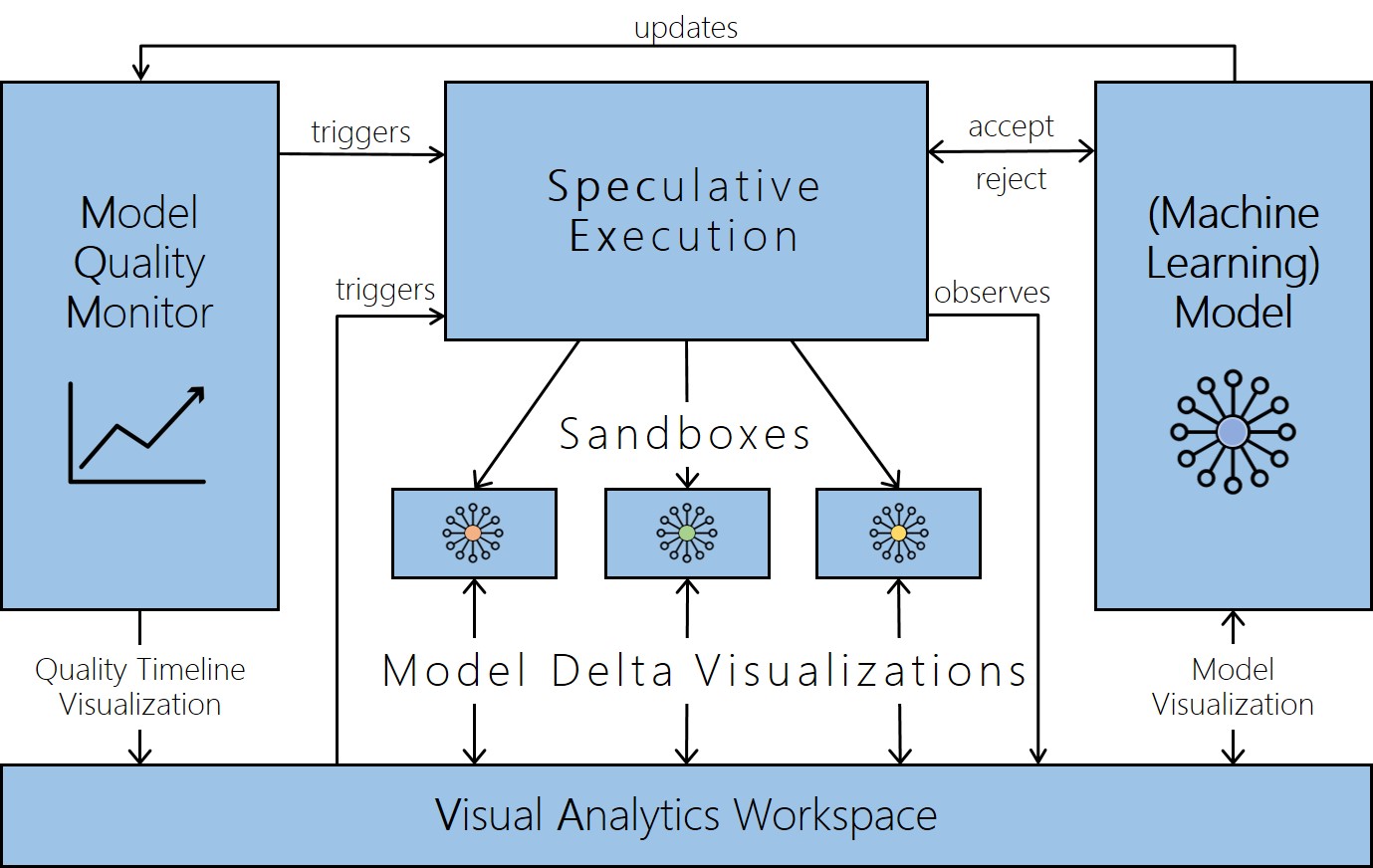}
\end{wrapfigure}
which is increasingly supported with VA techniques having the human in the loop in interactive machine learning settings.
In their current state, these VA systems support users in the labeling process with visual-interactive interfaces showing data characteristics as well as information about the current model state~\cite{BernardVialTVCJ}.
Example labeling interfaces include scatterplots~\cite{bernardTVCG2017} in combination with dimensionality reduction~\cite{sacha16}, radvis-like visualizations~\cite{SG10}, or list-based interfaces~\cite{eurova2018}, e.g., in combination with active learning models~\cite{settles2012}.
Interaction techniques that enable the assignment of labels are based on simple item selection or drag-and-drop facility.
Users of the systems start by creating a small set of training data with some labeling interactions, triggering the machine learning algorithm to be re-trained in an iterative way.
Intermediate results of the models are visualized in the labeling interfaces which closes the human-centered feedback loop~\cite{BernardVialTVCJ}.

Observing these interactions such a system could understand and learn the pattern in the users' interaction. 
It could then find potentially misclassified data instances by using quality metrics or assessing the spatial relationships of data instances in the high-dimensional data space or the visual space, or both.
In a series of parallel sandboxes, the system could ``auto-complete'' different interaction patterns the users have started, and present a list of changes together with the re-trained alternative models. The user can then inspect the list of instances which switched their prediction of the machine learning model and accept or reject the sandbox. 

If the system can detect and understand these \emph{semantic interactions}~\cite{endert2014human} as implicit steering commands, it can learn about the user's intuition (intents~\cite{yi2007}). It can then apply this intuition to the remaining data, removing the need for users to explicitly complete their interaction pattern, reassigning even more instances. 

\subsection{User Guidance}
Recent work in \acl{va} has often focussed on measuring 
\begin{wrapfigure}[7]{r}{0.4\columnwidth} 
    \vspace{-0.25em}
    \includegraphics[width=\linewidth]{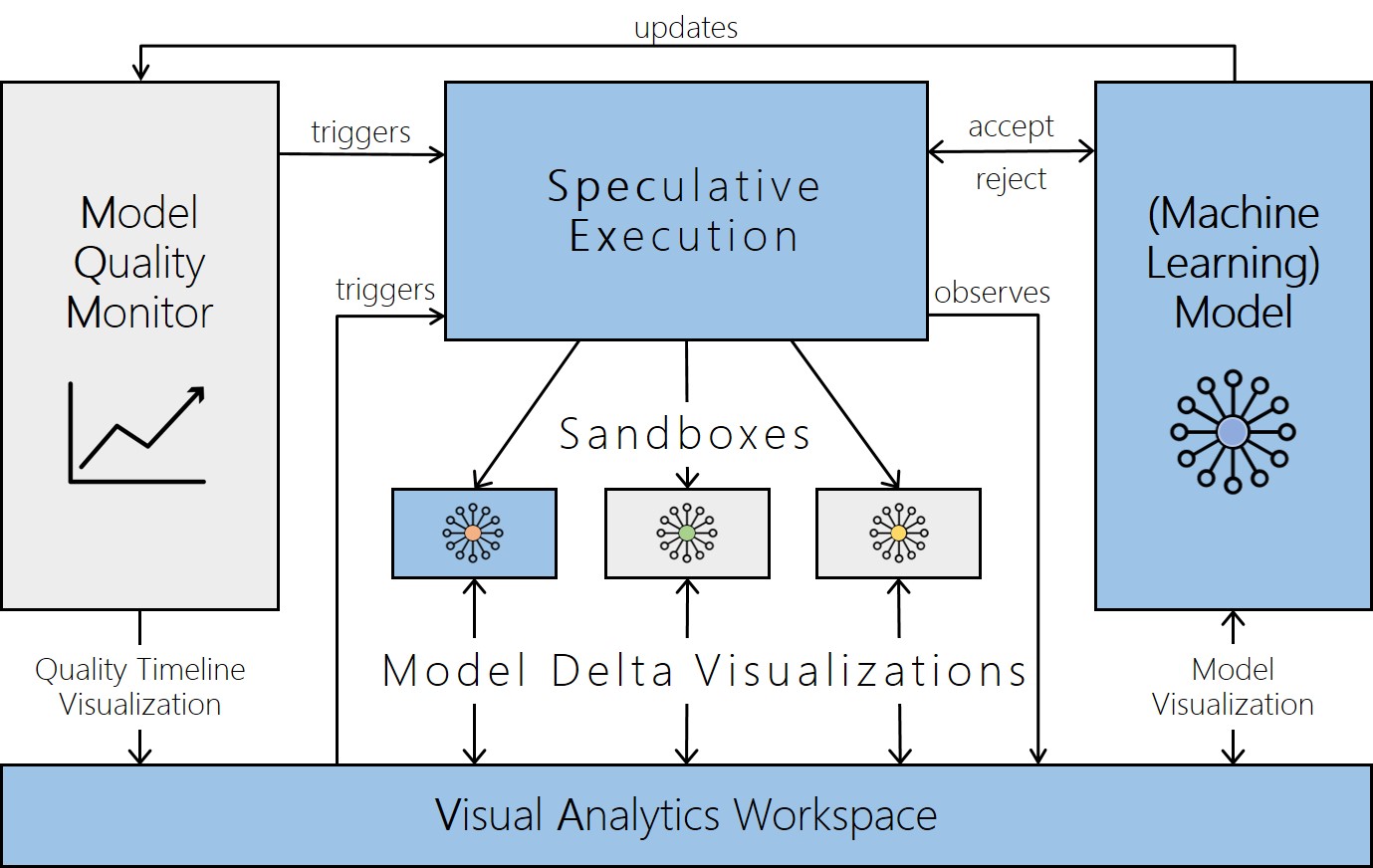}
\end{wrapfigure}
and avoiding confirmation bias~\cite{wall2017warning, arnott2006cognitive}. \acl{se} can help to prevent confirmation bias by providing alternative sandboxes, as well. A different approach has been taken by Wall et al. with their system PODIUM~\cite{wall2018podium}. It aims at making the users mental model visible by letting them rank data according to their intuition and preference. In their example with college football teams, users drag and drop some teams that they have an opinion about to a new position in the list. Additionally, users can indicate whether the model should put more or less weight on individual features. The system then learns feature weights from that ranking and reorders the list according to the learned model.  As a result, the relative differences between teams ranked by the user can change. Here, \acl{se} could explore alternative feature weightings that lead to a ranking that is closer to what the user originally expressed. 
The system could then guide users and highlight that changing certain feature weights would lead to the model more closely representing their originally expressed order. Users can then verify the changes and accept them, if they agree.
Alternatively, this could lead to users realizing that the proposed changes do not fit their mental model. As a result, they might start questioning the said model, exploring the data further, and overcoming their biases. This scenario shows that \ac{se} can not only be used to speed up model optimization or auto-complete user interactions. Instead, it can be used to explain model changes by highlighting features with a high impact on the current modeling situation.

\subsection{Cooperative Visual Analytics}
Linguists are interested in classifying questions on whether they are 
\begin{wrapfigure}[7]{r}{0.4\columnwidth} 
    \vspace{-0.25em}
    \includegraphics[width=\linewidth]{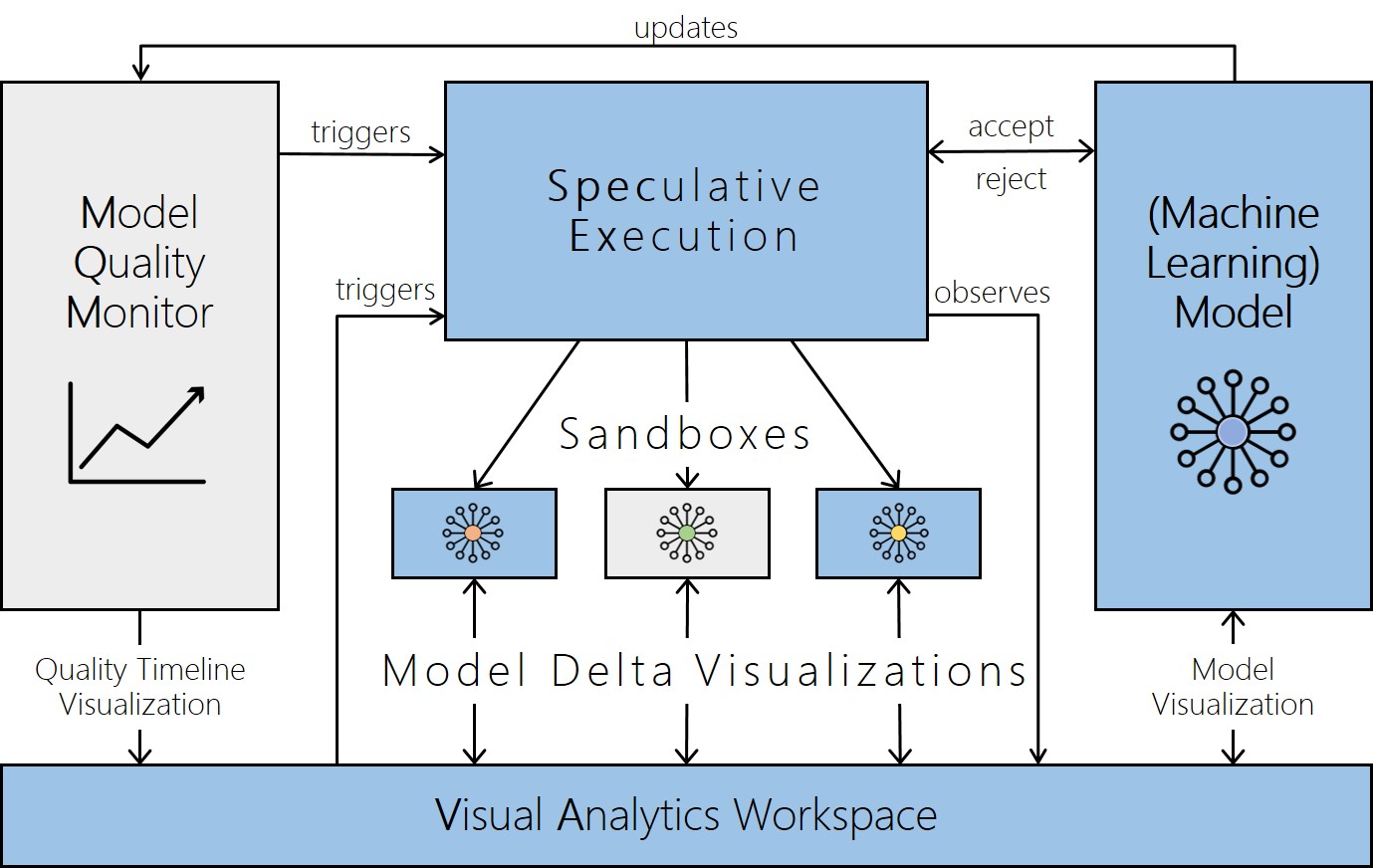}
\end{wrapfigure}
information-seeking (ISQ) or non-information-seeking (NISQ), i.e., rhetorical~\cite{sevastjanova_mixed_initiative}. Their data consists of transcripts of conversations or written text. The task of an expert analyst is to train a classifier using a \ac{va} system. Such a system might, for example, provide the context before and after the question, as well as information on the respective speakers. This task is interesting for cooperative analysis because discussions with linguists have shown that there is frequent disagreement on whether a question is an ISQ or not, even between experts. In current classifier training systems, this would lead to prolonged decisions on how to classify training data. With \acl{se}, disagreeing experts can create two sandboxes, and each classify some training data according to their understanding. If this task is executed on a large, collaborative touchscreen, both experts can even train their model in parallel. After retraining the respective models, the experts can compare which model captured the particularities of the current dataset better. In a provenance-tracking view, they can compare which training decisions were most important for making one of the classifiers more adapt to the data. After a short discussion, the experts can now agree on a common understanding of ISQ/NISQ that is suitable for the dataset. 

\subsection{Speculative Execution beyond Machine Learning}
Speculative Execution can also be extended beyond machine
\begin{wrapfigure}[7]{r}{0.4\columnwidth} 
    \vspace{-0.25em}
    \includegraphics[width=\linewidth]{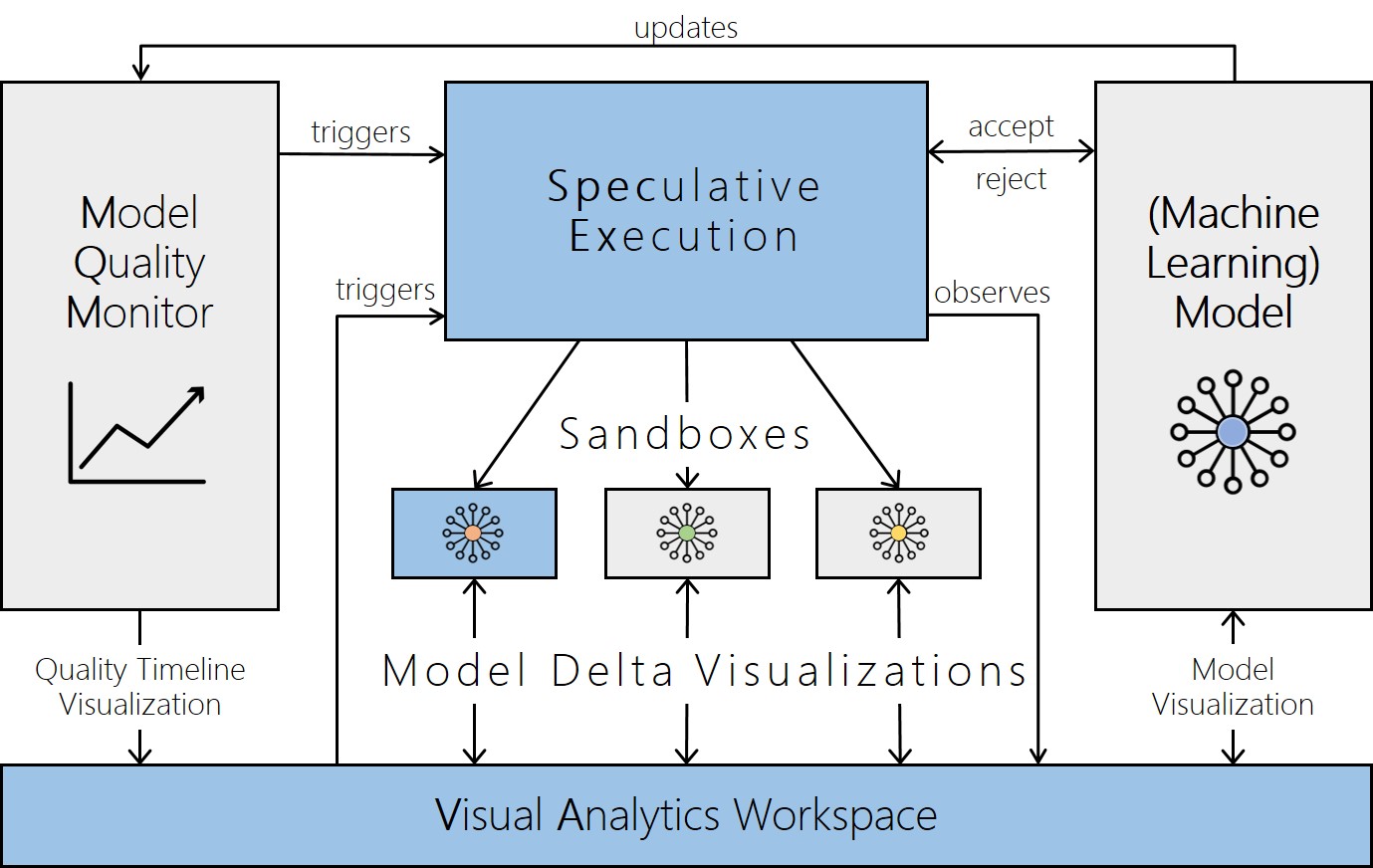}
\end{wrapfigure}
learning  scenarios. One possible extension is towards visualization systems that focus on presenting data without an underlying model. In this scenario, we use the example of a system visualizing geographical movement data of cars on a map. The ``envirocar'' data set\footnote{https://old.datahub.io/dataset/envirocar, last accessed 7/15/2018} contains GPS tracks of driving cars and is annotated with speed, the fuel consumption, rpm, et cetera. As it contains about 1.7 million rows, it cannot be visualized on a map without some preprocessing and aggregation. One such aggregation step is the bundling of trajectories to avoid overplotting. Additionally, a system may show a summary of the presented data in a detail panel. Such information is, for example, useful to families searching for a home in a quiet area without too much traffic and with low emissions. To make the interaction with the system more seamless, it uses eye-tracking to determine regions that users are interested in. Once such a region is determined, it zooms in and provides more detail. In combination with \ac{se} this zooming process is smoother. While the system is still determining whether a user is interested in a region or was only glimpsing at it, it can already precompute the aggregations on the new level and prepare the detail panel. If the zoom-in action is performed later, the data is already preprocessed and can be displayed. 

\section{Aspects of Speculative Execution}
Having shown the applicability of \ac{se} in various \ac{va} scenarios, we provide a list of theoretical considerations. They stem from both our experience with implementing \ac{se} for topic modeling and the hypothetical usage scenarios. We conclude by providing design principles for \ac{se}.}

\subsection{Semantic Complexity of Interactions}
\label{sec:speculation_levels}
User interaction can build the basis for relevant input for \ac{se}.
Depending on the level of \emph{semantic complexity} of such an interaction, the goals of the \acl{se} are different. We define semantic complexity as the amount by which the mental model of the user changes by performing this interaction. The different levels will be described in the following.

\paragraph{L1: Start of Interaction} Once the user has started an interaction, the system can assess what the user is trying to achieve. 
It can then try to predict how this interaction can be finished using extrapolation techniques. The goal of \acl{se} on this level is to prepare the context of the interaction target. During a drag and drop operation, for example, such computations could include the search for relevant ``drop targets''. 
\paragraph{L2: Completion of Interaction}
Every interaction that was completed can build a basis for similar interactions that may be performed in future.
Whenever an interaction has been finished, the system can try to predict the next interaction on a similar level of complexity that the user might want to perform. The goal of such speculations is to guide the user in exploring the potential impact of interactions. Also, \acl{se} at this level could make users aware of data points that might have been missed, for example when removing outlier nodes.
\paragraph{L3: Repetition of Interactions}
In many cases, users combine different low-level interactions to solve a higher-level intent~\cite{yi2007}.
As such, combinations of interactions are at the highest of the semantical levels of speculation that we want to address.
Tracking and contextualizing of low-level interactions to make sense of user intents builds the basis for the support of \ac{se} at this level. Once a user intent has been identified, the system can begin searching for similar intents that users might have, and that could be solved with a similar set of repeated low-level interactions.

\vspace{1em}
To some extent, the goal on all levels of semantic complexity is ``User Intention Guessing'': the system needs to determine the \emph{implicit interaction}~\cite{endert2014human} that the user is trying to achieve by performing the explicit interactions. Once the user's intention has been identified, appropriate optimizations, parameter changes, or additional computations can be searched in a much more focused manner. 

\subsection{Dimensions of Speculation}
\label{sec:speculation_dimensions}

\acl{se} and sandboxes provide the theoretical framework for an efficient, mixed-initiative guidance approach to visual analytics, extending the Visual Analytics pipeline by Keim et al.~\cite{Keim2008VisualChallenges}. Whenever \ac{se} is not proposing optimizations to complete user interactions, it can target improvement in the measured model quality and prepare alternative models. 
Their usefulness depends on the model-specific \emph{speculation dimensions}. These speculation dimensions are defined by parameters or properties of the underlying model. In any computed sandbox, some of the possible speculation dimensions may be altered with respect to the ``original'' model. The following section highlights guidelines for the selection of speculation dimensions. Although \ac{se} is not limited to those, we present five distinct categories of dimensions here.
\paragraph{Temporal Dimensions}
Temporal dimensions include the actual time as well as an iteration number, depending on the model. 
They are especially interesting for incremental algorithms. 
Incremental streaming models can build a buffer of events or data and ``forecast'' the development of the model in any given sandbox. For iterative algorithms, a sandbox can show the model development over the next iterations. 
\emph{Progressive} sandboxes can be continuously refined as new data becomes available. This can allow users to pursue multiple model alternatives in parallel, before deciding for one.
\paragraph{Optimization Strategies}
For various models, direct optimizations, manipulating the models' underlying data structures, can be conceived. Operations could include merging or splitting tree nodes, changing values of matrices and vectors, or introducing a threshold. These optimization strategies can be tailored towards known potential model issues and provide bespoke solutions for these problems, without having to explore and change the model's parameters. Such strategies have been implemented in the system presented in \ref{sec:model_optimization}~\cite{El-Assady2018SpecEx} and provide sandboxes avoiding topic chaining or combining small, overly specific topics into more easily understood generalized ones.

Utilizing bespoke optimization strategies incurs an additional implementation cost for identifying potential model issues and developing possible solutions. However, it can directly address problems that would be unintuitive, difficult, or even impossible to change through model parameter changes.

\paragraph{Model Parameters}
If no bespoke optimization strategies are available, the sandboxes of \ac{se} can perform a Visual Parameter Space Analysis~\cite{sedlmair2014visual}, creating different sandboxes for different parameter configurations. However, instead of sampling and precomputing the entire parameter space, the analysis can be focused on regions of the search space that are similar to the users current model. As soon as users start exploring new regions of the model state space, new sandboxes can be created and prepared in the background. 

\paragraph{Input Transformations}
If the preprocessing pipeline is integrated into the Visual Analytics system, \ac{se} sandboxes can utilize these preprocessing algorithms to transform the underlying data. Examples of such transformations include a stricter outlier-removal, filtering out stopwords from a collection of text documents, or introducing minimum and maximum-thresholds for time-series data. 

\paragraph{Algorithm Modifications}
Ultimately, \ac{se} sandboxes can also explore modifications of the original model or any of its parts. They can replace similarity functions, feature weighting schemes or merging strategies, to name just a few. As this dimension is specific to the underlying model, it can be very effective and powerful. As with any \emph{Input Transformations}, the system needs to explicitly inform the users of any changes made to these dimensions. Modifications here might have an impact on the user's mental model and how well it fits.

\subsection{Towards Design Principles}
\label{sec:design_principles}
In the following, we provide design considerations for effective \acl{se} in five areas.

\paragraph{Speculation Dimensions}
When performing speculative parameter space analysis or employing bespoke optimization strategies, there is a trade-off between the number of sandboxes that are created and their usefulness. As one goal of \acl{se} is a more guided exploration of the model state space, more sandboxes are beneficial: they lead to more model configurations being calculated and presented to the user.
However, in addition to the increased need for computational resources, users cannot and will not inspect and compare a large number of speculative sandboxes. In our first implementation of Speculative Execution for the optimization of topic models, we offered users the results of seven speculative optimization strategies~\cite{El-Assady2018SpecEx}. In the evaluation study, users often focused on the top-two or top-three optimizations according to our provided ranking. One possible reason is that model comparison is a difficult task. It might be alleviated by effective delta-visualizations, but users will remain unable to compare all sandboxes that can be computed. 

\paragraph{Runtime}
Depending on the level of semantic complexity of an interaction triggering speculative execution different runtime requirements apply. With increasing complexity of the performed interactions, increasingly complex Speculative Executions are necessary to support and guide the user. However, the runtime requirements for these more ``complex'' sandboxes triggered by L2 or even L3 interactions are not as strict as for those triggered by L1 interactions
that need to be executed while the user is performing an interaction like dragging and dropping an object. Here, the aim should be on focused and short computations of less than 500ms. As soon as the user ends the interaction, the speculation becomes meaningless. However, the resulting sandboxes for L2 and especially L3 interactions are likely still useful after a couple of seconds. It is important that such longer-running speculations do not block the user interface as to not interrupt the analysis workflow. 

\paragraph{Search Space Size}
\label{sec:search_space_size}
As we have previously implemented \ac{se} for IHTM~\cite{El-Assady2018SpecEx}, an incremental topic model building a tree structure, we elaborate on the size of the individual search- and model-spaces using a concrete example.
We consider a corpus containing 280 documents ($k$) and a two-dimensional speculative execution with $n=7$ optimization strategies and $b=10$ additional documents being inserted into the model from the buffer during a speculation. Running the model without optimization will produce exactly one topic-tree as its output. With \ac{se}, we compute at most $(k/b)\cdot n=196$ sandboxes. This allows us to involve the user in the algorithmic decision-making process but is significantly more scalable than considering all possible optimization paths (even considering the buffer), which would result in $n^{(k/b)} \approx 10^{23}$ options.  Even though this would be an exponentially large search space, it is still multiple orders of magnitude smaller than all possible trees an incremental algorithm would consider $(k! \approx 10^{565})$, or all possible trees with $k$ nodes $(k^{(k-2)} \approx 10^{680})$. 
This example shows that the \acl{se} reduces the factorial search space to a linear one. This is due to the use of only two dimensions with a very limited set of possible values. Here, careful considerations weighing search space exploration against computation time are necessary. One interesting area for future research is the formalization of a cost-benefit model for \acl{se} that can be used as guidance when selecting speculation dimensions. 

\paragraph{Quality Metrics}
Whenever \acl{se} is not triggered by user interaction it needs quality metrics to trigger and assess sandboxes. The user typically performs a multi-objective optimization of these metrics when trying to improve a model. As a result, good consensus strategies between the used metrics are necessary to sort the computed sandboxes before presenting them to the user for exploration. Note that it is typically not possible to fully automate this optimization process, as most quality metrics do not capture (all) semantic details.
\paragraph{Delta-Visualization}
For an efficient \ac{se}, an effective delta-visualization is necessary to highlight the differences between the current model and a selected sandbox. This visualization needs to be tailored to the underlying model, the visualization from which the \acl{se} was triggered, and the number of changes introduced in the speculation. Gleicher et al.~\cite{gleicher2011visual} introduced various patterns for visual comparison. While different patterns are useful in different situations, we argue that explicit encoding should be present in comparative sandbox visualizations to help users to quickly focus on the introduced changes that decide over accepting or rejecting the proposed sandbox. For some tasks like labeling, a list of elements with changed labels might be more useful than a complex visualization trying to highlight the differences between two dimensionality-reduction results.

\subsection{Research Opportunities}
With \acl{se} being a novel concept in \acl{va}, many interesting questions remain.

\paragraph{Mapping Interaction to Optimization}
Understanding the intent of user interactions is paramount for effective \acl{se}. While systems using implicit steering exist today, it is an open field of research how interactions on the different levels of semantic complexity can be understood and mapped to concrete goals for a speculative sandbox. 
Endert et al. have already identified the capturing of user interaction intentions as relevant future work~\cite{Endert15Toward}. In addition, they have elaborated on design considerations and confidence levels of captured interactions~\cite{Endert:2012:SIV:2207676.2207741}. Further research should investigate how such captured interactions can be generalized and re-applied to complete the user's semantic interactions.

\paragraph{Cost-Benefit-Model}
As was alluded to in the previous section, choosing the wrong (number of) speculation dimensions is detrimental to the usefulness of \ac{se}. An information-theoretic cost-model for \acl{se} could define the size of the search space, the number of visited states, the computation time and the cognitive load on users. Such a cost-model would then allow informed choices on the sandboxes \ac{se} computes. 

\paragraph{Interaction Design}
\acl{se} constantly computes alternative models that users might want to explore and needs to present them for inspection. However, systems should not constantly interrupt the analyst's workflow for model-comparison. Further research should be conducted to determine when and how to show sandbox results, and when to refrain from interrupting the user. Furthermore, research should investigate how different presentation styles of \ac{se} results impact the creation and confirmation of biases.

\section{Conclusion}

We have introduced \acl{se} as a new methodology for \acl{va}. 
Based on a definition and formalization of \ac{se} for \ac{va}, we presented five possible usage scenarios that demonstrate the applicability of \ac{se} in \ac{va}.
Finally, we characterized different aspects of \ac{se} in \ac{va}, showing that \ac{se} is a multifaceted concept that can support \ac{va} in different ways.
The primary benefits of successfully adopting \ac{se} into the \ac{va} processes are multiple ways in which user guidance can be provided algorithmically, conflated with visual-interactive interfaces.
Next steps include the implementation of \acl{se} in several \ac{va} applications to further demonstrate the applicability of \ac{se} for guided \ac{va}. Likewise, the elaboration of formerly discussed research opportunities will lead to new insights.

\acknowledgments{This work has been funded in part by the Deutsche Forschungsgemeinschaft (DFG) within the project ``Visual Analytics and Linguistics for Interpreting Deliberative Argumentation (VALIDA)'', Grant Number 376714276, as part of the Priority Program ``Robust Argumentation Machines (RATIO)'' (SPP-1999).}

\bibliographystyle{abbrv-doi-hyperref}

\bibliography{template}
\end{document}